\documentclass[aps,prx,superscriptaddress,twocolumn]{revtex4-2}

\usepackage{siunitx}
\usepackage{amsmath}
\usepackage{graphicx}
\usepackage{longtable}

\begin{document}
\title{A terahertz vibrational molecular clock with systematic uncertainty at the $10^{-14}$ level}

\author{K. H. Leung}
\thanks{These authors contributed equally to this work.}
\affiliation{Department of Physics, Columbia University, 538 West 120th Street, New York, NY 10027-5255, USA}
\author{B. Iritani}
\thanks{These authors contributed equally to this work.}
\affiliation{Department of Physics, Columbia University, 538 West 120th Street, New York, NY 10027-5255, USA}
\author{E. Tiberi}
\affiliation{Department of Physics, Columbia University, 538 West 120th Street, New York, NY 10027-5255, USA}
\author{I. Majewska}
\altaffiliation{Present address:  Procter and Gamble Polska Ltd.,
Zabraniecka 20,
03-872 Warsaw,
Poland}
\affiliation{Quantum Chemistry Laboratory, Department of Chemistry,
University of Warsaw, Pasteura 1, 02-093 Warsaw, Poland}
\author{M. Borkowski}
\affiliation{Department of Physics, Columbia University, 538 West 120th Street, New York, NY 10027-5255, USA}
\affiliation{Institute of Physics, University of Amsterdam, Science Park 904, 1098XH Amsterdam, The Netherlands}
\affiliation{Institute of Physics, Faculty of Physics, Astronomy and Informatics, Nicolaus Copernicus University, Grudziadzka 5, 87-100 Torun, Poland}
\author{R. Moszynski}
\affiliation{Quantum Chemistry Laboratory, Department of Chemistry,
University of Warsaw, Pasteura 1, 02-093 Warsaw, Poland}
\author{T. Zelevinsky}
\email{tanya.zelevinsky@columbia.edu}
\affiliation{Department of Physics, Columbia University, 538 West 120th Street, New York, NY 10027-5255, USA}

\date{\today}

\begin{abstract}
Neutral quantum absorbers in optical lattices have emerged as a leading platform for achieving clocks with exquisite spectroscopic resolution. However, the class of absorbers and studies of systematic shifts in these clocks have so far been limited to atoms. Here, we extend this architecture to an ensemble of diatomic molecules and experimentally realize an accurate lattice clock based on pure molecular vibration. We evaluate the leading systematics, including the characterization of nonlinear trap-induced light shifts, achieving a total systematic uncertainty of $4.6\times10^{-14}$. The absolute frequency of the vibrational splitting is measured to be 31 825 183 207 592.8(5.1) Hz, enabling the dissociation energy of our molecule to be determined with record accuracy. Our results represent an important milestone in molecular spectroscopy, THz frequency standards, and may be generalized to other neutral molecular species with applications for fundamental physics, including tests of molecular quantum electrodynamics and the search for new interactions.
\end{abstract} 

\maketitle

\section{Introduction}

The pursuit of high performance quantum clocks has historically spurred important developments, including laser cooling \cite{chu1985three,aspect1988laser,lett1988observation} and optical trapping \cite{ashkin1970acceleration,ashkin1978trapping}. In one highly successful clock architecture, large numbers of quantum absorbers are tightly confined in a magic wavelength optical lattice, affording reduced quantum projection noise \cite{katori2003ultrastable}. Such lattice clocks, so far employing atomic optical transitions, have realized record performance in both precision \cite{bothwell2022resolving,zheng2022differential,oelker2019demonstration,schioppo2017ultrastable,mcgrew2018atomic,bloom2014optical} and accuracy \cite{mcgrew2018atomic,bloom2014optical,bothwell2019jila,nicholson2015systematic,ohmae2021transportable,ushijima2015cryogenic,nemitz2016frequency,yamanaka2015frequency}, ushering in a new era in space-time sensing \cite{takamoto2020test,delva2017test,lisdat2016clock,origlia2018towards}. In parallel, there is growing interest in advancing the laser spectroscopy and quantum control of more complex particles\textbf{---}such as diatomic or polyatomic molecules with rich rovibrational structure\textbf{---}motivated by fundamental physics applications \cite{safronova2018search,mitra2022quantum} that include the search for particles beyond the Standard Model \cite{Andreev2018,cairncross2017precision,alauze2021ultracold,grasdijk2021centrex,hutzler2020polyatomic,yu2021probing}, fifth forces \cite{germann2021three,salumbides2013bounds,Borkowski2019scirep}, the time variation of the electron-to-proton mass ratio \cite{Kobayashi2019,hanneke2020optical,barontini2022measuring,zelevinsky2008precision,shelkovnikov2008stability}, and dark matter \cite{oswald2022search,kozyryev2021enhanced}. Molecules also hold promise as new platforms for quantum computation and simulation \cite{kaufman2021quantum,burchesky2021rotational,wang2022enriching,zhang2022quantum,albert2020robust,tesch2002quantum,altman2021quantum,moses2017new}, and the adoption of clock-inspired techniques may help extend the available coherence times for these applications. 

Recent experiments have shown quantum logic rotational spectroscopy of trapped $\mathrm{CaH}^+$ ions \cite{collopy2022rotational} and magic wavelength vibrational spectroscopy of $\mathrm{Sr}_2$ \cite{Kondov2019} at a precision of parts in $10^{13}$, as well as cavity ring-down rovibrational spectroscopy of buffer-gas cooled acetylene at few parts in $10^{12}$ \cite{aiello2022absolute}. However, accurate vibrational spectroscopy of neutral molecules at (or below) the $10^{-13}$ level remains difficult and unexplored \cite{patra2020proton,kortunov2021proton,molony2016measurement,cheng2018dissociation,beyer2019determination,hussels2022improved}.

Here, we extend the lattice clock architecture to trapped neutral molecules and characterize the systematic frequency shifts in a $\mathrm{Sr}_2$ vibrational lattice clock, achieving a total fractional systematic uncertainty of $4.6\times10^{-14}$, comparable to the earliest realizations of optical atomic lattice clocks \cite{takamoto2005optical,ludlow2006systematic,le2006accurate}. We measure the pure molecular vibrational frequency to 13 digits, establishing one of the most accurately known oscillator frequencies in the terahertz (THz) band thus far \cite{riehle2018cipm,shelkovnikov2008stability}. We leverage this to characterize the electronic ground potential of the strontium dimer\textbf{---}originating from the van der Waals bonding of two closed-shell atoms\textbf{---}by determining the dissociation energy of $^{88}\mathrm{Sr}_2$ with an accuracy surpassing the previous record for a diatomic molecule \cite{molony2016measurement}. 

The results described here may be applied to a wide class of molecules (e.g., hydrogen isotopologues \cite{Jozwiak2022}), enabling the refinement of molecular quantum electrodynamics calculations, tests of fundamental laws, and potentially open new pathways for THz frequency metrology \cite{tonouchi2007cutting,Wang2018,nagano2021terahertz}.

\section{Vibrational Clock \label{sec:methods}}

\begin{figure*}
\centering
\includegraphics[width=2\columnwidth]{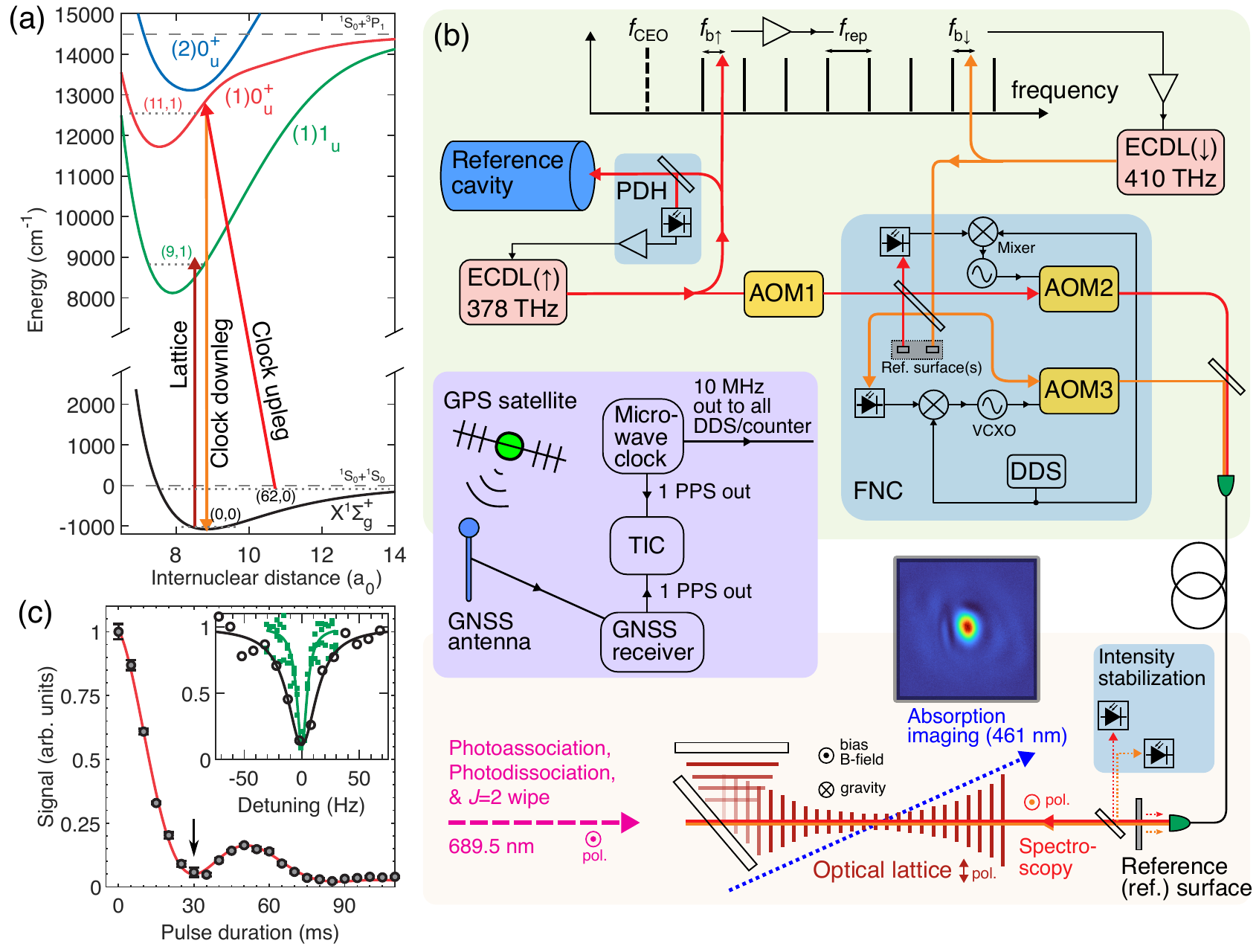}
\caption{Vibrational molecular lattice clock.  (a) Raman lasers (upleg, red arrow; downleg, orange arrow) detuned from an intermediate state in $(1)0_u^+$ probe the two-photon vibrational clock transition between $(v=62, J=0)$ and $(v=0, J=0)$ in the $X^1\Sigma_g^+$ ground potential. The optical lattice (brown arrow) off-resonantly addresses an isolated rovibronic state in $(1)1_u$ to induce magic trapping conditions. (b) Experimental setup. The upleg master laser is stabilized to a reference cavity using the Pound-Drever-Hall (PDH) technique, and its phase coherence is transferred to the downleg laser via a frequency comb. The molecules are held in the 1D optical lattice. Co-propagating clock lasers are delivered to the molecules via an optical fiber with active fiber noise cancellation (FNC). The spectroscopic signal derives from absorption imaging of $X(62,0)$ photofragments at a slight grazing angle relative to the lattice. A Rb microwave standard acts as a flywheel oscillator, linking the molecular clock to GPS time for the absolute frequency measurement. Further information is given in the main text and Appendices~\ref{sec:atomicprep} and \ref{sec:ramanlaser}. (c) Two-photon Rabi oscillations between the clock states driven at the operational probe intensities (filled circles, experimental data averaged over 8 consecutive runs, error bars represent $1\sigma$ uncertainties; solid red line, analytical fit to an exponentially decaying sinusoid). We observe lines as narrow as 11(1) Hz (inset, green squares). For clock operation, we perform Rabi spectroscopy with a 30 ms $\pi$-pulse duration (indicated by the black arrow), resolving 30(2) Hz linewidths consistent with the expected Fourier limit (inset, black open circles). Each point in the inset is a single shot of the experiment, and solid lines are Lorentzian fits.}
\label{fig:expscheme}
\end{figure*}

The basic scheme of the molecular clock is as follows. We operate the clock on the pure vibrational transition between the weakest bound and most tightly bound irrotational states, $(v=62, J=0)\rightarrow(v=0, J=0)$, in the $X^1\Sigma_g^+$ ground potential of $^{88}\mathrm{Sr}_2$. Here, $v$ and $J$ denote the vibrational and total angular momentum quantum numbers, respectively. This pair of clock states offers the largest possible pure vibrational splitting (or clock frequency, $f_\mathrm{clock}\sim32 \,\mathrm{THz}$) in the ground state for our molecule, and can be used to obtain the molecular dissociation energy ($D_0$) given the binding energy of $X(62,0)$. As a direct transition between $J=0$ states is strictly forbidden, we drive the clock transition via a Raman process using two diode lasers detuned from the intermediate excited state $(1)0_u^+(v'=11,J'=1)$. This pathway through a deeply bound $(1)0_u^+$ state offers favorable Rabi frequencies, which facilitated stimulated Raman adiabatic passage (STIRAP) transfer between $X(62,0)$ to $X(0,0)$ in our preceding work \cite{leung2021ultracold}. By contrast, weakly bound $(1)0_u^+$ states near the intercombination line (which we utilized in Ref.~\cite{Kondov2019}) are expected to have negligibly small transition strengths to $X(0,0)$ due to poor Franck-Condon overlap \cite{leung2020transition,Skomorowski2012pra}. The relevant potentials are shown in Fig.~\ref{fig:expscheme}(a). 

The measurements take place in a retroreflected one-dimensional (1D) optical lattice at $\sim$1005 nm. Trapped samples of ultracold molecules are created by photoassociating laser cooled strontium atoms at 2 $\mu$K to the $(1)1_u(v'=-1,J'=1)$ rovibronic state. This efficiently produces $X(62,0)$ ground state molecules thanks to the large transition strength \cite{leung2020transition}. 
Molecules formed in the undesired $J=2$ excited rotational state are photodissociated, and the remaining atoms are wiped out of the trap with resonant 461 nm laser light. Our detection scheme relies on state-selective photodissociation of $X(62,0)$ followed by absorption imaging of the slow-moving atoms. As this destroys the molecular sample, the entire sequence is iterated to scan the clock transition. Appendix \ref{sec:atomicprep} contains an elaboration on the state preparation.

Raman clock spectroscopy is deeply in the Lamb-Dicke regime for co-propagating probes along the axial direction of the optical lattice (Lamb-Dicke parameter $\eta_\mathrm{LD} \lesssim 0.02$). The upleg (or pump) master clock laser at 378 THz (793 nm) is stabilized to a high finesse ultra-low expansion reference cavity with a measured drift rate of $30 \,\mathrm{mHz}/\mathrm{s}$ that is compensated using a linearly-ramped acousto-optic modulator. The phase coherence of the upleg is transferred to the teeth of an erbium-fiber-laser-based optical frequency comb by actuating on its repetition frequency [Fig.~\ref{fig:expscheme}(b)]. The downleg (or anti-Stokes) clock laser at 410 THz (731 nm) is phase locked to the comb, thereby inheriting the phase stability of the upleg. The carrier-envelope offset frequency of the comb is stabilized to a Rb standard that serves as the laboratory timebase. Since the Raman transition samples the correlated frequency difference of the clock lasers, spectral broadening due to the linewidth of the upleg master laser is greatly suppressed. The upleg is passed through an acousto-optic modulator (AOM1 in Fig.~\ref{fig:expscheme}(b)), whose first order diffraction is used to iteratively step the difference frequency of the clock lasers across $f_\mathrm{clock}$. AOM1 controls the interrogation duration by pulsing the upleg, and we leave the downleg constantly irradiating (but blocked with a mechanical shutter during the state preparation process). Both clock lasers are delivered to the molecules via the same optical fiber, and active fiber noise cancellation \cite{ma1994delivering,rauf2018phase} is implemented separately on each leg using the same phase reference surface; see Appendix \ref{sec:ramanlaser}.

Figure~\ref{fig:expscheme}(c) shows two-photon Rabi oscillations driven by the clock lasers at the operational Rabi frequencies. Using pulse durations of $\sim$100 ms, our apparatus is capable of producing clock lines with full width at half-maximum as narrow as 11(1) Hz, corresponding to a $Q$-factor of $2.9 \times 10^{12}$ (solid green squares in the inset to Fig.~\ref{fig:expscheme}(c)). In the absence of other fields, the blackbody radiation (BBR) limited lifetimes of the clock states exceed $10^5$ years in a room temperature environment \cite{Kondov2019}, suggesting no fundamental limit for the $Q$-factor. Nevertheless, the main technical challenges in the current iteration of the molecular clock are two-body molecular losses close to the universal rate \cite{leung2021ultracold} and lattice light-induced one-body losses for $X(0,0)$ that scale quadratically with the trap depth \cite{Kondov2019,leung2020transition}. At the operational density and lattice trap depth in this work, these losses quench the spectroscopic signal fast enough that the molecular densities vary significantly over pulse durations of $\gtrsim$60 ms. As a compromise, we evaluate clock systematics by performing Rabi spectroscopy with a 30 ms $\pi$-pulse, scanning Fourier-limited peaks of 30(2) Hz (black open circles in the inset to Fig.~\ref{fig:expscheme}(c)). A typical spectrum consists of 15 experimental iterations (taking a total duration of $\sim$20 s), from which we determine the line center by fitting a Lorentzian function.

\section{Results}
\subsection{Systematic evaluation}
Table \ref{tab:systable} details the uncertainty budget of the molecular clock under the operational conditions of this work. Summing the uncertainties of all contributors in quadrature, we report a total systematic uncertainty of $4.6\times 10^{-14}$.

We leverage the short-term frequency stability of our reference cavity to average down the uncertainty of a given systematic. Most frequency corrections in Table \ref{tab:systable} are determined by probing the clock transition in an interleaved fashion; i.e., we alternate an experimental parameter between two values and record the corresponding pair of line centers \footnote{For the density shift measurements, the molecule numbers were changed cycle-to-cycle (i.e., interlaced). For the probe and lattice light shift measurements, the laser intensities were changed scan-to-scan, due to the speed at which the motorized neutral density filters could reliably switch positions. To further suppress the effect of residual cavity drift not canceled by the feedforward acousto-optic modulator, we employ three-point string analysis \cite{nicholson2015systematic}, which simulates simultaneous interrogation. This has the drawback of artificially correlating and approximately doubling the number of frequency shifts extracted from the dataset. Hence, we multiply the standard error of the weighted mean by approximately $\sqrt{2}$ to extrapolate back to the expected level of statistics had we used independent, non-overlapping, successive pairs of interleaved measurements to extract the frequency shifts. This is in addition to scaling up the standard error by the square root of the reduced chi-square statistic if it exceeds 1 to give the revised (or scale-corrected) standard error, which accounts for statistical overscatter.}. This is repeated to gather statistics, and the shift in the line center is computed as a weighted average. The clock shift, $\Delta f_\mathrm{clock}$, is defined as the frequency shift relative to the unperturbed clock. The clock shifts are extrapolated to determine the frequency correction for the clock at the operational parameter value. We quote the revised (or scale-corrected) standard errors for all fit parameters obtained from weighted fitting or averaging; i.e., we scale up the statistical uncertainties of the fit parameters by the square root of the reduced chi-square statistic ($\chi^2_\mathrm{red}$) if $\chi^2_\mathrm{red} > 1$, which indicates an overscattered dataset relative to the fitted model.

\begin{table}
 \caption{Systematic uncertainty budget for the strontium molecular clock under operating conditions. See Appendix \ref{sec:othersys} for the description of minor systematics not in the main text. All values are expressed in fractional units ($\times 10^{-14}$).}
    \label{tab:systable}
\begin{ruledtabular}
    \centering
    \begin{tabular}{lcc}
    Systematic & Correction& Uncertainty\\
    \colrule
    Lattice Stark ($E1,M1,E2$) &100.1&3.4\\
    Lattice Stark (hyperpolarizability) &-50.8&1.9\\
    Probe Stark (total) &31.5&2.2\\
    BBR &-2.2&0.4\\
    Density&-0.6&0.3\\
    Quadratic Zeeman& 0&0.05\\
    dc  Stark & 0&$<0.1$\\
    Doppler and phase chirps& 0& $<1$\\
    Lattice tunneling& 0&$<0.1$\\
    Line pulling& 0&$<0.1$ \\
    Scan-and-fit& 0&$<0.6$\\
    \colrule
    \textbf{Total}&77.9&4.6
    \end{tabular}
    \end{ruledtabular}
\end{table}

\subsubsection{Lattice light shift}

Magic\textbf{---}or state-insensitive\textbf{---}trapping conditions can be engineered for the vibrational clock states by off-resonantly addressing $X(0,0)\rightarrow (1)1_u(9,1)$ with the lattice. In the present work, the protocol involves predominantly tuning the polarizability of $X(0,0)$ to match that of $X(62,0)$ at a magic detuning of 4.493(3) GHz \cite{leung2021ultracold}. Owing to a favorably large Franck-Condon factor, the transition strength between $X(0,0)\rightarrow (1)1_u(9,1)$ is the largest among all $X(v,0)\rightarrow (1)1_u(v',1)$ transitions \cite{leung2021ultracold,leung2020transition}, resulting in a magic detuning that is nearly $4\times$ larger than that in our initial demonstration of the molecular magic wavelength technique \cite{Kondov2019}, putting less stringent requirements on the frequency stability of the lattice laser. 

Importantly, the neighboring $(1)1_u(v',1)$ rovibronic resonances are spaced at intervals of $\sim$2 THz, and may cause deleterious shifts due to lattice light impurity (e.g., amplified spontaneous emission (ASE) \cite{fasano2021characterization}). To mitigate this, the lattice light derives from a Ti:sapphire laser stabilized to the same optical frequency comb described in Section \ref{sec:methods}. This also permits the lattice frequency, $f_\mathrm{latt} = c/\lambda_\mathrm{latt}$, to be determined with kHz-level accuracy. To further suppress ASE impurity at the magic detuning, the light is filtered through a linear cavity (finesse of 50, and free spectral range of 2.9 GHz) before delivery to the experiment by a single-mode polarization maintaining fiber. A stable, weak reflection from the vacuum window is used for lattice intensity stabilization during normal operation. The linear lattice polarization defines the quantization axis for the magnetically insensitive $X^1\Sigma_g^+$ states \cite{mcguyer2015control}. 

\begin{figure}
\centering
\includegraphics[width=\columnwidth]{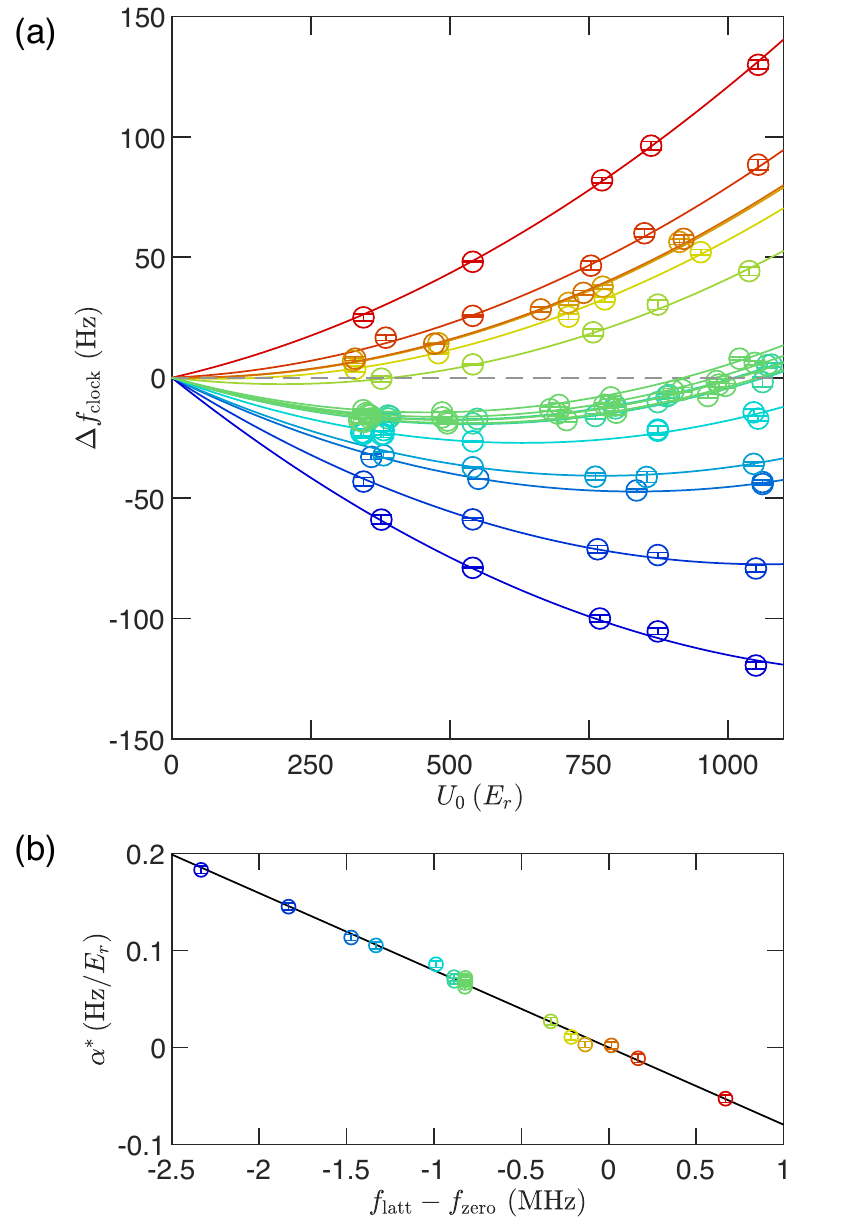}
\caption{(Color online) Clock shifts due to the lattice light.  (a) Nonlinear shifts of the molecular clock frequency versus trap depth. For a given lattice frequency (color coded), we make interleaved measurements of clock shifts (open circles) with respect to a reference trap depth ($\sim500\,E_r$), and fit the data to parabolas (solid lines) with a global quadratic parameter, $-\beta^*$. (b) Linear light shift coefficient, $\alpha^*$, versus lattice frequency (color code matches (a)), and the linear fit (black solid line). $\alpha^*$ is predominantly due to the $E1$ differential polarizability and is nulled at $f_\mathrm{zero}$. By tuning $\alpha^*$, we can find conditions where the sensitivity of $\Delta f_\mathrm{clock}$ to fluctuations in $U_0$ is minimal at our operational trap depth of 487(4) $E_r$ (dark green points). Error bars represent 1$\sigma$ uncertainties.}
\label{fig:latticelightshifts}
\end{figure}

We investigate the effect of lattice light over a range of $f_\mathrm{latt}$. At each $f_\mathrm{latt}$ we make interleaved measurements of the clock shifts, alternating the trap depth $U_0$ between a reference depth and four other depths spanning from 300 $E_r$ to 1100 $E_r$, where $E_r \equiv h^2/(2M\lambda_\mathrm{latt}^2)$ is the recoil energy and $M$ is the molecular mass. The trap depths are determined from the axial trapping frequencies (Appendix \ref{sec:trapcal}). Small corrections ($<0.3\times10^{-14} \times f_\mathrm{clock}$) were made to account for density shifts. As shown in Fig.~\ref{fig:latticelightshifts}(a), our measurements reveal nonlinear light shifts as a consequence of molecular hyperpolarizability. The higher-order transitions that account for this effect will be investigated in future work, but we hypothesize a connection with previously observed quadratic lattice scattering rates in a similar experiment \cite{Kondov2019}.

In order to characterize the lattice light shifts, we adopt the thermal model described in Ref.~\cite{brown2017hyperpolarizability} and write the clock shifts as \begin{equation}\label{eq:lattshifteqn}
    \Delta f_\mathrm{clock} = -\alpha^*U_0 - \beta^*U_0^2,
\end{equation} where $\alpha^*$ and $\beta^*$ are empirically obtained from parabolic fits to the measured differential shifts. These parameters are effective values dependent on the trapping conditions: $\alpha^*$ is related to the differential electric-dipole ($E1$), magnetic-dipole ($M1$) and electric-quadrupole ($E2$) polarizabilities, while $\beta^*$ is related to the differential hyperpolarizability. Crucially, the polynomial form of Eq.~(\ref{eq:lattshifteqn}) hinges on a linear scaling of the sample temperature with $U_0$, which we verify to hold true for our molecules using Raman carrier thermometry (Appendix \ref{sec:trapcal}). We do not expect non-polynomial terms \cite{UshijimaKatori2018} to be significant at the level of the current evaluation.

The fits give $\beta^* = -6.81(22) \times 10^{-5}\,\mathrm{Hz}/E_r^2$ as a global parameter. Additionally, the results for $\alpha^*$ versus $f_\mathrm{latt}$ are shown in Fig.~\ref{fig:latticelightshifts}(b), and a linear fit yields a sensitivity slope $\partial \alpha^*/\partial f_\mathrm{latt} = -0.0796(16) \,\mathrm{Hz}/(\mathrm{MHz}\, E_r)$ as well as an $x$-intercept $f_\mathrm{zero} = 298\,368\,568.844(21)\,\mathrm{MHz}$. Operating the molecular clock at a trap depth of $U_\mathrm{opt} = 487(4)\,E_r$ and $f_\mathrm{latt}-f_\mathrm{zero} = -0.821(21)\,\mathrm{MHz}$, we determine the correction terms to be $\alpha^*U_\mathrm{opt} = 31.8(1.1)\,\mathrm{Hz}$ and $\beta^*U_\mathrm{opt}^2 = -16.2(6)\,\mathrm{Hz}$, summing to a fractional correction of $49.3(3.8)\times 10^{-14}$. Under these conditions, $\Delta f_\mathrm{clock}$ is first-order insensitive to changes in $U_0$ (dark green points in Fig.~\ref{fig:latticelightshifts}).

\subsubsection{Probe light shift}

\begin{figure}
\centering
\includegraphics[width=\columnwidth]{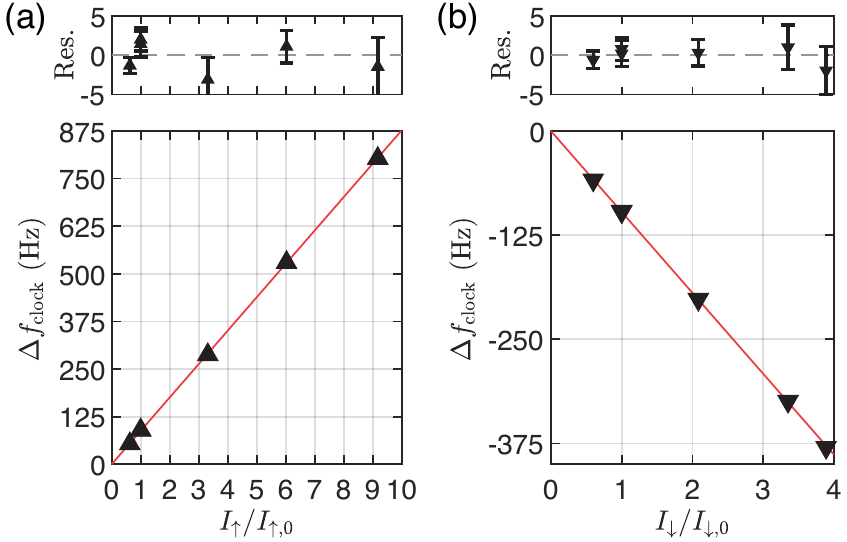}
\caption{Clock shifts at the operational Raman detuning as a function of (a) the upleg laser intensity, and (b) the downleg laser intensity. The horizontal axes are normalized by the respective operational intensities, $I_{\uparrow,0}$ and $I_{\downarrow,0}$. Solid lines are linear fits to the data. Residuals are plotted in units of Hz. Error bars represent 1$\sigma$ uncertainties.}
\label{fig:probelightshift}
\end{figure}

Probe light shifts pose an inherent challenge for two-photon spectroscopy. This is even more so for scalar clock states ($J=0$) that preclude the use of laser polarization-based cancellation schemes \cite{jackson2019magic}. Here, the clock shifts scale linearly as the probe intensities are low, and are related to the differential polarizability at the respective probe wavelength ($\lambda_p$), \begin{equation}\label{eq:probeshifteqn}
\Delta f_\mathrm{clock} = \frac{I_{p}}{2h\epsilon_0c}\left[\alpha_0(\lambda_{p})-\alpha_{62}(\lambda_{p})\right], \end{equation} where $\alpha_v$ is the $E1$ polarizability for the vibrational state $v$, $I_{p}$ is the probe laser intensity, and $p\in\{\uparrow,\downarrow\}$ specifies the laser: upleg ($\uparrow$) or downleg ($\downarrow$). Figure~\ref{fig:probelightshift} shows that linear extrapolation of probe shifts suffices for a molecular clock at the few $10^{-14}$ level.

While tailored pulse sequences to alleviate probe light shifts have been proposed \cite{yudin2018generalized,zanon2016probe,zanon2006cancellation,hobson2016modified}, for this evaluation we opted for a more straightforward strategy. We can minimize the total probe light shift by using so-called balanced intensity ratios satisfying the condition $I_{\uparrow}\left[\alpha_0(\lambda_\uparrow)-\alpha_{62}(\lambda_\uparrow)\right]  = -I_{\downarrow} \left[\alpha_0(\lambda_\downarrow)-\alpha_{62}(\lambda_\downarrow)\right]$. At the same time, a large Raman detuning\textbf{---}relative to the intermediate $(1)0_u^+(11,1)$ excited state\textbf{---}is preferred so that off-resonant scattering from the probes have a negligible effect on the accessible coherence times. Figure~\ref{fig:probelightshift} demonstrates that such conditions exist in our clock for blue detunings where the baseline polarizability differences at the probe wavelengths have opposite signs, in agreement with our polarizability model (Appendix \ref{sec:abinitiopol}). We operate at a Raman detuning of +14.973 GHz, much greater than the 5 MHz natural linewidth of the intermediate state \cite{leung2021ultracold}.

We evaluate $\Delta f_\mathrm{clock}$ for each leg separately. Using a motorized neutral density filter, we switch between two intensity values for one leg while keeping that of the other leg at its operational value. The $\pi$-pulse durations are adjusted accordingly. Typical settings for the interleaved measurements are $(P_{\uparrow,0},9 P_{\uparrow,0})$, and $(P_{\downarrow,0},3.5 P_{\downarrow,0})$ where $P_{p,0}=I_{p,0}(\pi w_p^2/2)$ are the operational powers measured with a calibrated power meter immediately before the vacuum window. These shifts are scaled by the measurement lever arms to obtain the clock corrections at the operational settings: $-(\Delta f_\mathrm{clock}/\Delta P_p) \times P_{p,0}$. We find the corrections to be $-277.5(1.4)\times 10^{-14}$ for the upleg, and $309.0(1.7)\times 10^{-14}$ for the downleg. Drifts in $\Delta P_{p}$ are at the sub-percent level over the $\sim$2000 s duration for each probe light shift evaluation, and the weighted averages of $f_\mathrm{clock}$ typically have $\chi^2_\mathrm{red}\sim1$. Accurate knowledge of the beam waists $w_p$ is not necessary as they are robust during an evaluation, and they are common factors that drop out in calculations. Long-term drifts due to beam pointing instability may be monitored and countered by benchmarking the probe intensities using the molecules (e.g., using an Autler-Townes frequency splitting, an on-resonance scattering rate, or the two-photon Rabi oscillation frequency), which we leave to future work.

\subsubsection{Blackbody radiation shift}

Homonuclear dimers are infrared inactive, conferring natural insensitivity to blackbody radiation (BBR). Using the formulas derived in Ref.~\cite{porsev2006multipolar}, the frequency correction due to BBR is calculated to be $-0.70(14)\,\mathrm{Hz}$ at an effective temperature of $T^\mathrm{eff} = 303(5)\,\mathrm{K}$; see Appendix \ref{sec:chambertherm} for a description of the chamber thermometry. The uncertainty is dominated by \textit{ab intio} calculations of the dc  polarizabilities of the clock states (Appendix \ref{sec:abinitiopol}). Comparison with experimentally measured ac polarizability ratios show agreement at the level of 10--20\%, to be expected from typical accuracies of theoretical transition strengths. Therefore, we assign a conservative fractional uncertainty of 20\% for the BBR shift.

\subsubsection{Density shift}

\begin{figure}
\centering
\includegraphics[width=\columnwidth]{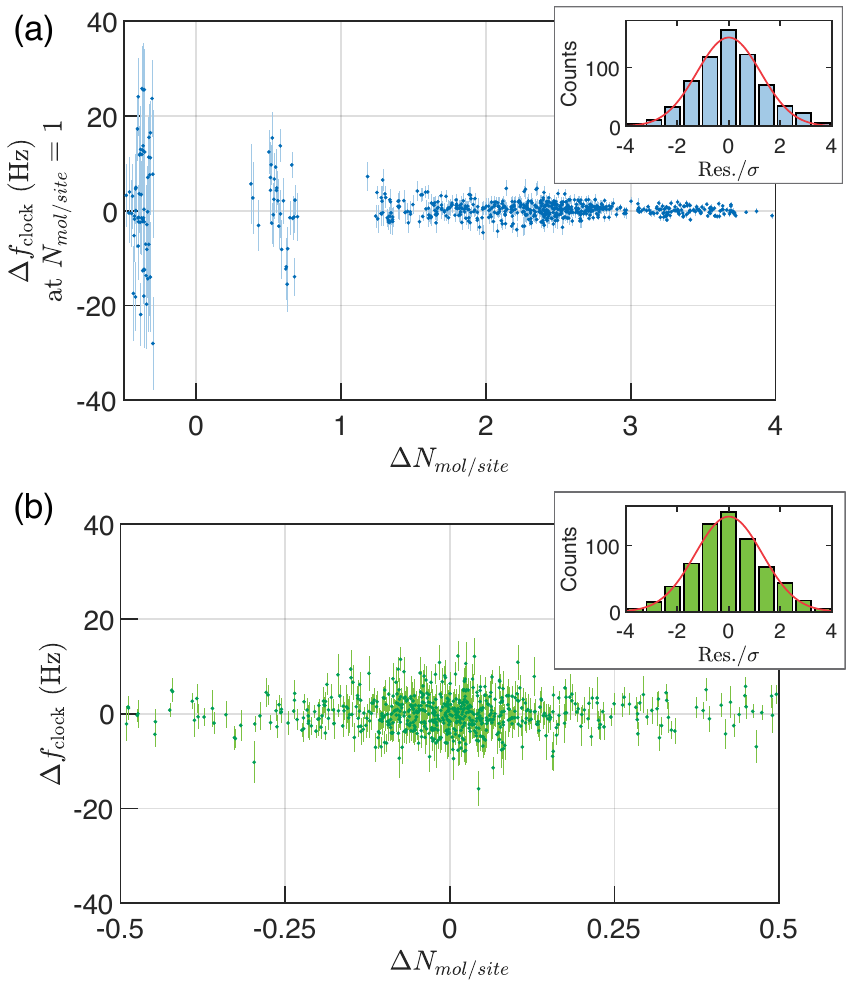}
\caption{Density shift evaluation.  (a) Clock shifts due to molecular collisions extrapolated to operating conditions (1 molecule per lattice site, averaged over filled sites), plotted versus the change in molecule number per site used for the interleaved measurement. A single constant suffices to fit the data (0.20(10) Hz, $\chi^2_\mathrm{red}=1.7$). (b) In the same dataset, the shift between successive resonances taken under identical experimental settings serves as a control experiment to check for technical offsets. As expected, this averages to zero (0.03(20) Hz, $\chi^2_\mathrm{red}=2.0$). All statistical errors are scaled up by $\sqrt{\chi^2_\mathrm{red}}$. Error bars represent 1$\sigma$ uncertainties. Both insets show the histogram of normalized residuals, and the solid red lines are Gaussian fits. }
\label{fig:densityshift}
\end{figure}

Due to their bosonic character, our $^{88}\mathrm{Sr}_2$ molecules are unprotected against $s$-wave collisions. The one-dimensional lattice forms a series of microtraps, each with a trap volume proportional to $(T/
\tilde{\omega}^2)^{3/2}$. Here, $T$ is the temperature of the molecules, and $\tilde{\omega}$ is the geometric mean of the angular trapping frequencies. We investigate density dependent shifts arising from dimer-dimer collisions by modulating the average number of molecules per lattice site ($N_\mathrm{mol/site}$) at the beginning of the clock pulse. This is achieved by inserting a wait time immediately after photoassociation (PA) so that two-body collisions naturally reduce the molecule number \cite{leung2021ultracold,leung2020transition}. Fluctuations in $N_\mathrm{mol/site}$ are typically $<20\%$, and we assume equal occupancy across filled sites. Since both $T$ and $\tilde{\omega}^2$ scale similarly with $U_0$, and the lattice intensity is stabilized, $N_\mathrm{mol/site}$ is a robust observable proportional to the molecular density. 

Assuming linear density shifts, we scale our differential measurements to find $\Delta f_\mathrm{clock}$ at the normal operating value of $N_\mathrm{mol/site}=1$. Figure~\ref{fig:densityshift}(a) summarizes the measurements performed at various number differences ($\Delta N_\mathrm{mol/site}$) suggesting a correction of $-0.20(10)\,\mathrm{Hz}$, or $-0.63(31)\times10^{-14}$ in fractional units, due to collisional shifts. Control measurements using spectra taken under common experimental settings do not show evidence of spurious offsets in our data [Fig.~\ref{fig:densityshift}(b)].  It is instructive to compare the size of our density shift with similarly performing atomic clocks. From a trap calibration (Appendix \ref{sec:trapcal}) we estimate that the shift coefficient has a magnitude of $2.9(1.5)\times10^{-25}\,\mathrm{cm^{3}}$ after normalizing by the transition frequency. This is rather similar to the analogous optical atomic clock with bosonic $^{88}\mathrm{Sr}$ ($\sim 2\times10^{-25} \,\mathrm{cm^{3}}$ \cite{lisdat2009collisional}), while being orders of magnitude smaller than in Cs ($\sim1\times10^{-21}\,\mathrm{cm^{3}}$ \cite{gibble1993laser,dos2002controlling}) or Rb microwave clocks ($\sim5\times10^{-23}\,\mathrm{cm^{3}}$ \cite{sortais2000cold}). 

\subsection{Absolute frequency evaluation}

As illustrated in Fig.~\ref{fig:expscheme}(b), we reference all RF frequency counters and direct digital frequency synthesizers (DDS) in the experiment to a free-running Rb microwave standard (our local timebase). Calibration of this Rb clock is accomplished by comparing its 1 pulse-per-second (PPS) output with that of a dual-band global navigation satellite system (GNSS) receiver using a time interval counter (TIC); see Appendix~\ref{sec:timebase}. The Rb clock serves as a flywheel oscillator to access Global Positioning System (GPS) time. 

Each measurement trial of the absolute clock frequency is performed under operational conditions, where the molecular clock systematics are controlled at the level quoted in Table~\ref{tab:systable}. We repeatedly scan the clock transition to obtain a time series of the line centers, while simultaneously counting the repetition rate of the frequency comb. The probe light shifts were evaluated every trial to account for potential daily variations in probe laser beam pointing. The correction due to gravitational redshift is given in Appendix~\ref{sec:othersys}.

Figure~\ref{fig:absfreq} shows the results of the measurement campaign, consisting of 10 trials performed on separate days. A weighted average yields the absolute frequency of the $^{88}\mathrm{Sr}_2$ vibrational clock to be $f_\mathrm{clock} =$ 31 825 183 207 592.8(5.1) Hz, with a fractional uncertainty of $1.6\times 10^{-13}$.


\begin{figure}
\centering
\includegraphics[width=\columnwidth]{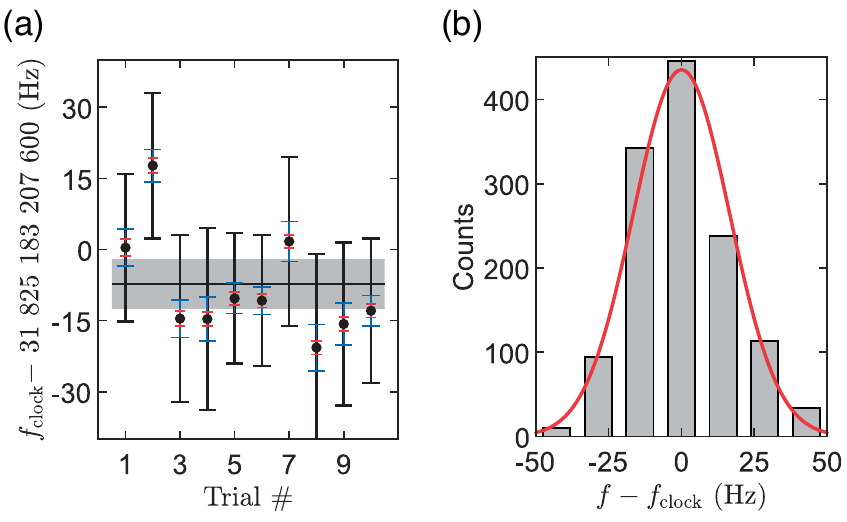}
\caption{(a) Absolute frequency of the clock transition measured over 10 trials (filled black circles) with all known frequency offsets corrected, including that of the local Rb timebase (see main text for details). Blue error bars are 1$\sigma$ statistical uncertainties, dominated by the determination of the comb repetition rate rather than the stability of the scanned molecular clock lines. Red error bars are 1$\sigma$ systematic uncertainties due to the molecular clock only (see Table~\ref{tab:systable}). Black error bars are 1$\sigma$ total uncertainties, where the uncertainties of the local timebase calibrations are added in quadrature with the statistical and molecular clock systematic uncertainties. The black horizontal line shows the weighted average ($\chi^2_\mathrm{red} = 0.5$), and the shaded grey area shows the associated $\pm1\sigma$ standard error of the mean. (b) Histogram of all clock frequency measurements in the 10 trials, relative to the weighted average of $f_\mathrm{clock}$. The solid red line is a Gaussian fit to the histogram.}
\label{fig:absfreq}
\end{figure}

\section{Conclusion}

Few frequency standards currently exist in the THz band \cite{shelkovnikov2008stability,riehle2018cipm}. Our molecular clock serves as a THz reference and can generate stable radiation at 9.4 $\mu$m via photomixing \cite{preu2011tunable,hindle2011widely}. Alternatively, transitions in heteronuclear isotopologues could be driven directly with quantum cascade lasers \cite{bartalini2014frequency,consolino2019qcl}. To our knowledge, $f_\mathrm{clock}$ represents one of the most accurately measured pure molecular vibrational frequencies to date. The fractional uncertainty is on par with that of the unidentified rovibrational interval in $\mathrm{OsO}_4$ near the $\mathrm{R}(10)\,(00^01)\text{--}(10^00)$ emission line of the $\mathrm{^{12}C^{16}O}_2$ laser. This absorption line in $\mathrm{OsO}_4$ is a secondary representation of the SI second \cite{riehle2018cipm}, and was compared directly against a primary cesium standard by stabilizing a $\mathrm{CO}_2$ laser to the specific saturated absorption feature of $\mathrm{OsO}_4$ in a high-finesse cavity \cite{daussy2000performances,rovera2001optical}. We expect to reduce the uncertainty of our local timebase calibration to the same level as the molecular clock systematics (or better) by upgrading to a standard with intrinsically lower instability and utilizing two-way time transfer schemes. 

Molecular spectroscopy is increasingly appreciated as a fertile ground in the search for new physics. The reported Hz-level molecular clock is a starting point for elucidating the bonding of the $\mathrm{Sr}_2$ dimer across a large range of internuclear distances. The isotopologues of $\mathrm{Sr}_2$ have different nucleon numbers, and comparison of their vibrational spectra may permit the investigation of hypothesized hadron-hadron interactions \cite{salumbides2013bounds}. 

The sum of $f_\mathrm{clock}$ with the binding energy of the least bound state $X(62,0)$ yields the dissociation energy ($D_0$) of our molecule with respect to the ${^1S}_0+{^1S}_0$ threshold. While the analogous least bound vibrational states of $^{84}\mathrm{Sr}_2$ and $^{86}\mathrm{Sr}_2$ are known with sub-kHz uncertainties \cite{stellmer2012creation,aman2018photoassociative}, the current best measurement for $^{88}\mathrm{Sr}_2$ is at the kHz-level \cite{McDonald2017}. Nevertheless, taking the binding energy of $X(62,0)$ to be $136.6447(50) \,\mathrm{MHz}$ from Ref.~\cite{McDonald2017}, which was determined using two-photon dissociation, we find $D_0(^{88}\mathrm{Sr}_2) = \text{31 825 319 852(5) kHz}$, or $\text{1 061.578 402 09(17)} \,\mathrm{cm}^{-1}$. This is an improvement by 5 orders of magnitude over the previously reported value for $\mathrm{Sr}_2$ in available literature \cite{Stein2010}, and sets a new accuracy record for the determination of a molecular dissociation energy ($1.6\times 10^{-10}$ fractional uncertainty). To list a few competitive results, dissociation energies have been reported with fractional uncertainties of $4.4\times 10^{-10}$ for $^{87}\mathrm{Rb}^{133}\mathrm{Cs}$ \cite{molony2016measurement}, $6.9\times 10^{-10}$ for ortho-$\mathrm{H}_2$ \cite{cheng2018dissociation}, $8.6\times 10^{-10}$ for para-$\mathrm{H}_2$ \cite{beyer2019determination}, and $7.1\times 10^{-10}$ for ortho-$\mathrm{D}_2$ \cite{hussels2022improved}.

Gaining access to longer coherence times is a general strategy for improving the systematic uncertainty. It would enable the excitation of narrower molecular resonances, expediting the evaluation of a systematic shift. Operating at lower trap depths would considerably suppress the lattice light-induced one-body losses of the deeply bound vibrational state, $X(0,0)$. To this end, we plan to reorient the lattice in a future upgrade such that its tighter axial dimension is along gravity to permit the confinement of molecules in a shallower trap. Notably, atomic lattice clocks have entered the $\sim 10 \,E_r$ regime \cite{kim2022evaluation}, and adopting these recent techniques should further mitigate the lattice light-induced losses. This may be supplemented by deeper atomic cooling \cite{zhang2022sub,akatsuka2021three} prior to photoassociation. Moreover, given that the lattice light shift is the most significant systematic in this work, operation at shallower trap depths would directly improve the clock accuracy. Longer Rabi interrogation times imply smaller effective Rabi frequencies, thus the operational probe laser intensities can be reduced, which lowers the total probe light shift. Future work may circumvent collisional shifts and losses by preparing samples with single molecule occupancy in a three-dimensional lattice \cite{akatsuka2008optical,akatsuka2010three,takano2017precise,kato2012observation}, or an optical tweezer array \cite{zhang2022optical,yu2021coherent,madjarov2019atomic,young2020half}.

In summary, we have demonstrated a vibrational molecular clock with a total systematic uncertainty of $4.6\times10^{-14}$, entering a new domain in high-resolution molecular spectroscopy. Our results are enabled by merging the key strengths of atomic clock techniques with molecular quantum science. 

\begin{acknowledgments}
We gratefully thank J. Sherman and T. H. Yoon for advice on the absolute frequency measurement and insightful discussions, and V. Lochab for early contributions to the vacuum chamber thermometry. We also thank the anonymous referees for their careful reading of the manuscript and useful suggestions. This work was supported by NSF grant PHY-1911959, AFOSR MURI FA9550-21-1-0069, ONR grant N00014-21-1-2644, a Center for Fundamental Physics grant from the John Templeton Foundation \& Northwestern University, the Brown Science Foundation, and the Polish National Science Centre (NCN) grant 2016/20/W/ST4/00314.  M. B. was partially funded by the Polish National Agency for Academic Exchange within the Bekker Programme, project PPN/BEK/2020/1/00306/U/00001, and by NCN, grant 2017/25/B/ST4/01486.
\end{acknowledgments}

\appendix

\section{State preparation \label{sec:atomicprep}}
Our experiments start with a thermal beam of $^{88}\mathrm{Sr}$ atoms decelerated by a Zeeman slower and laser cooled in a first-stage (blue) magneto-optical trap (MOT) using the ${^1S}_0\rightarrow {^1P}_1$ transition at 461 nm. We repump on the ${^3P}_2\rightarrow {^3S}_1$ and ${^3P}_0\rightarrow {^3S}_1$ transitions at 707 nm and 679 nm respectively. A second-stage (red) MOT on the narrower ${^1S}_0\rightarrow {^3P}_1$ intercombination at 689 nm further cools the atoms to a typical temperature of $2 \,\mu\mathrm{K}$. Throughout the cooling sequence ($\sim 500$ ms), a one-dimensional optical lattice at 1005 nm overlaps with the atom cloud, and atoms with kinetic energies lower than the trap depth are loaded into the lattice. The surrounding magnetic field is lowered to $<$0.6 G to prepare for molecule production and spectroscopy. The lattice is formed by retroreflecting the lattice laser beam and, in this study, oriented horizontally with respect to gravity due to practical limitations. 

Spin statistics and molecular symmetry imply that only even values of total angular momentum $J$ exist for ground state $^{88}\mathrm{Sr}_2$ molecules. While rotational factors favor the decay from $1_u(J'=1)$ to $X(J=0)$, a finite number of $X(J=2)$ molecules still form. Thus, to purify the gas, we photodissociate the $X(62,2)$ molecules 30 MHz above the ${^1S}_0+{^3P}_1$ threshold, imparting more than sufficient kinetic energy to guarantee these photofragments leave the trap. We do this concurrently with photoassociation (PA) by adding a frequency sideband to the PA laser with an acousto-optic modulator. We use a PA pulse duration of $\sim$2 ms. The remaining atoms are blasted out of the trap with resonant 461 nm laser light. For the operational trap depth in this study, we prepare $6\times 10^3$ molecules in the initial clock state $X(62,0)$, spread across approximately 520 lattice sites. To mitigate density shifts during clock operation, we further hold the molecules for a short duration ($\sim$150 ms), leveraging on the natural two-body inelastic collisions to reduce lattice occupancy to 1 molecule per site, averaged over filled lattice sites. Finally, photodissociation of $X(62,0)$ near the ${^1S}_0+{^3P}_1$ threshold produces slow-moving atoms that we absorption image to use as our spectroscopic signal [Fig.~\ref{fig:expscheme}(b)]. Mechanical shutters provide secondary shuttering of all laser beams (except the lattice) prior to entering the vacuum chamber, in addition to the fast primary shuttering performed using acousto-optic modulators. In this manner, we ensure that the molecules interact only with the Raman clock lasers and the lattice during clock interrogation. A complete account of our molecule production and detection methods can be found in Refs.~\cite{Reinaudi2012,McGuyer2015high}. 

\section{Raman clock lasers \label{sec:ramanlaser}}

Our reference cavity is formed from two fused silica mirrors bonded to an ultra-low expansion glass spacer placed in a vacuum housing maintained at the measured zero-crossing temperature for the coefficient of thermal expansion. The cavity finesse is $>3\times 10^5$ from ring-down measurements. The upleg diode laser serves as the master clock laser, and is stabilized to the cavity using the Pound-Drever Hall technique. We phase lock the repetition rate of an erbium-fiber-laser-based optical frequency comb directly to the upleg. Observations of the counted repetition rate against a Rb standard actively steered by a GPS disciplined oscillator for over a month prior to the campaign reveal a cavity drift rate of $30 \,\mathrm{mHz}/\mathrm{s}$, which we compensate using an acousto-optic modulator in the optical path of the master laser to the cavity. The frequency synthesizer that performs this linear feedforward compensation updates every second. The residual linear drift of the master laser due to imperfect feedforward is approximately $3 \,\mathrm{mHz/s}$ during the campaign, consistent with the observed drift of the molecular clock line centers over the same period after accounting for the comb teeth difference ($3 \,\mathrm{mHz/s} \times [1-(731 \,\mathrm{nm})/(793\,\mathrm{nm})] \approx 0.2 \,\mathrm{mHz/s}$). By phase locking the downleg diode laser directly to the comb, the comb acts as a transfer oscillator, and the phase fluctuations of the clock lasers are correlated. To suppress phase fluctuations due to the microwave synthesizers, the beats of the clock lasers with the comb are chosen to have the same sign and frequency; i.e., $f_\mathrm{b\uparrow}=f_\mathrm{b\downarrow}$ in Fig.~\ref{fig:expscheme}(b).

The clock lasers are injected into the same polarization-maintaining single-mode fiber and delivered to the adjacent optical table where the experiments take place. Since the wavelengths are sufficiently different that the laser beams may sample non-identical paths in a given refractive medium, active fiber noise cancellation (FNC) on each clock leg is implemented using independent phase actuators (acousto-optic modulators AOM2 and AOM3 in Fig.~\ref{fig:expscheme}(b)). The voltage-controlled crystal oscillators (VCXO) provide the RF frequencies for AOM2 and AOM3, and the FNC beats are phase locked to the same RF reference derived from a direct digital frequency synthesizer (DDS). The phase reference surface at the experiment table is a single partially reflecting mirror, while the surfaces on the laser table are mounted on a common rigid pedestal post with the clock lasers approaching the surfaces in the same direction. To minimize the number of optical elements and unstabilized path lengths, the clock lasers interrogate the molecules from the opposite direction as the PA and photodissociation lasers. The total uncompensated path in air is approximately 50 cm. The polarizations of the probes are identical, linear, and parallel to the small applied magnetic field, but perpendicular to that of the lattice in this work. During the state preparation sequence described in Appendix~\ref{sec:atomicprep}, both clock lasers are blocked by a mechanical shutter before the beams enter the vacuum chamber. The $1/e^2$ beam waists of the upleg and downleg laser beams at the molecules are 89(5) $\mu$m and 114(20) $\mu$m, respectively.

\section{Trap calibration and Raman carrier thermometry \label{sec:trapcal}}

\begin{figure}
\centering
\includegraphics[width=\columnwidth]{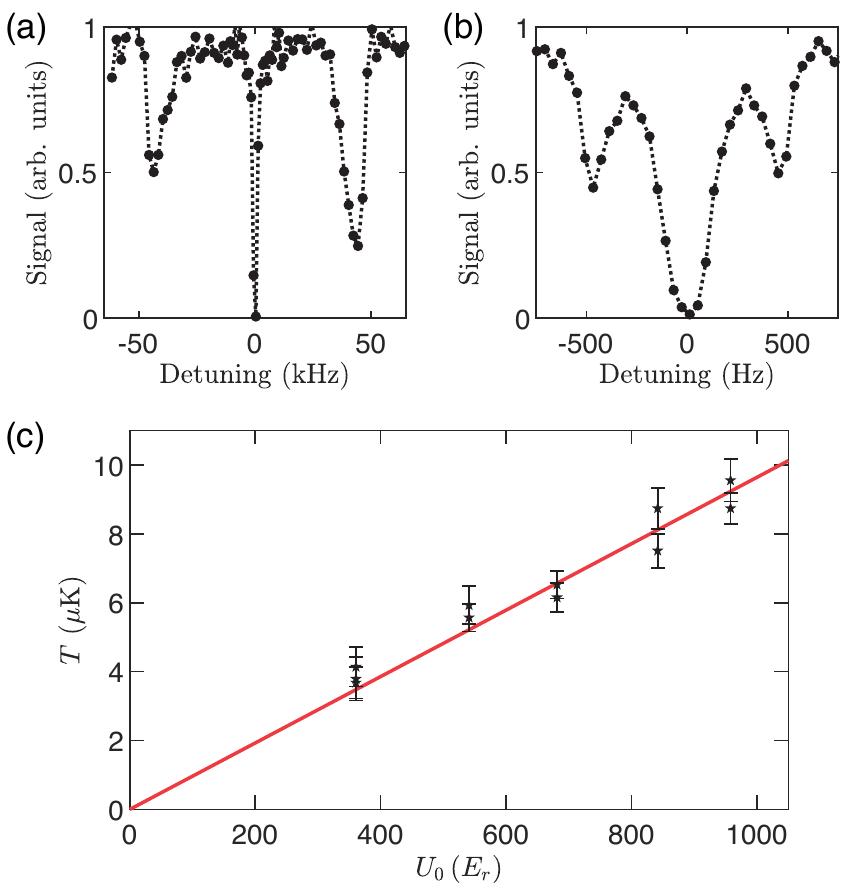}
\caption{Exemplary spectra of molecular (a) axial and (b) radial sidebands excited using Raman clock transitions, from which the motional frequencies are extracted to calibrate the lattice trap. (c) Measurement of the molecular temperature as a function of trap depth (black stars) using Raman carrier thermometry. Error bars denote $1\sigma$ uncertainties. A linear fit (solid red line) with the intercept fixed at the origin describes the data well ($\chi^2_\mathrm{red} = 0.92$).  \label{fig:trapcalapp}}
\end{figure}
The axial ($f_\mathrm{ax}$) and radial ($f_\mathrm{rad}$) trap frequencies for the molecules are measured with resolved-sideband spectroscopy at the operational magic lattice wavelength [Figs.~\ref{fig:trapcalapp}(a,b)]. To enhance the transition rates for the axial sidebands, we use counter-propagating probes to interrogate the naturally nearly-magic Raman transition between the adjacently bound vibrational states $X(62,0) \rightarrow X(61,0)$. To excite the radial sidebands, we use the Raman clock transition $X(62,0)\rightarrow X(0,0)$ but intentionally introduce a small relative misalignment in the probe beams. The trap depths are calculated using $U_0 = M\lambda^2f_\mathrm{ax}^2/2$. We note that, in general, $U_0/E_r = M^2\lambda_\mathrm{latt}^4f_\mathrm{ax}^2/h^2$, scaling as the square of the particle mass. Thus given the same trap frequencies and $\lambda_\mathrm{latt}$, $U_0/E_r$ is numerically $4\times$ larger for $\mathrm{Sr}_2$ molecules than $\mathrm{Sr}$ atoms. Assuming a Boltzmann thermal distribution, the microtrap volumes are calculated as $V = \left[ 2\pi k_BT/(\tilde{\omega}^2M) \right]^{3/2}$, where $\tilde{\omega}\equiv 2\pi(f_\mathrm{ax}f_\mathrm{rad}^2)^{1/3}$ and $M$ is the molecular mass. 

To determine $T$, the temperature of the molecular ensemble, we perform Raman carrier thermometry \cite{mcdonaldzelevinsky2015,leung2021ultracold} with co-propagating probes scanning $X(62,0)\rightarrow X(0,0)$ at a non-magic wavelength (tuned $>$0.3 THz away from the operational magic wavelength to maximize the polarizability difference). As shown in Fig.~\ref{fig:trapcalapp}(c), the dependence of $T$ against $U_0$ is well described by a linear fit to the data. 

Lastly, the ratio $f_\mathrm{ax}^2/f_\mathrm{rad}^2$ taken at the same trap laser intensity is related to the $1/e^2$ beam waist ($w_\mathrm{latt}$) of the lattice at its focus; i.e., $f_\mathrm{ax}^2/f_\mathrm{rad}^2 = w_\mathrm{latt}^2\,(2\pi/\lambda_\mathrm{latt})^2/2$. We find $w_\mathrm{latt} = 36(1) \,\mu\mathrm{m}$.

\section{Vacuum chamber thermometry \label{sec:chambertherm}}

Short of finite-element modeling, we may make a basic estimate for the effective solid angle ($\Omega^\mathrm{eff}_\mathrm{angle,i}$) subtended by the $i$-th surface surrounding the molecular sample as \begin{equation}
    \frac{\Omega^\mathrm{eff}_\mathrm{angle,i}}{4\pi} = \frac{\Omega_\mathrm{angle,i} \eta_i}{\sum_i \Omega_\mathrm{angle,i} \eta_i},
\end{equation} where $\eta_i$ is the surface emissivity and $\Omega_\mathrm{angle,i}$ is the geometric solid angle. The total effective solid angle is normalized to $4\pi$. We use values for the emissivity of various materials from available literature \cite{bothwell2019jila,handbook2009american,barnes1947total,wieting1979effects,wittenberg1965total}. These are 0.91 for glass (fused silica), 0.54 for sapphire, and 0.08 for stainless steel. 

The sapphire window facing the Zeeman slower is heated to 430(10) K and subtends a geometric solid angle of 0.04 sr. Among the fused silica window viewports with direct line-of-sight to the molecules, there are 8 with a diameter of 33.78 mm, 4 with a diameter of 69.85 mm, and 6 with a diameter of 114.3 mm. The diameter of the spherical vacuum chamber is approximately 240(10) mm, and the surface area consisting of stainless steel is approximated as the spherical surface area subtracted by the total area encompassed by the viewports. 

At the present level of precision, it is enough to estimate the temperature environment of the stainless steel vacuum chamber using four negative temperature coefficient (NTC) thermistors affixed to its exterior. The largest (smallest) sensor reading is $T_\mathrm{c,max}$ $(T_\mathrm{c,min})$. We model the temperature gradient as a rectangular distribution \cite{joint2008evaluation} and estimate the temperature of the vacuum chamber to be $(T_\mathrm{c,max}+T_\mathrm{c,min})/2=302\,\mathrm{K}$, with an uncertainty of $(T_\mathrm{c,max}-T_\mathrm{c,min})/\sqrt{12}=1\,\mathrm{K}$. Conservatively, the fused silica windows are within $\pm$2 K of the temperature of the stainless steel chamber. The line of sight from the molecules to the hot oven is blocked using an in-vacuum mechanical shutter during clock spectroscopy. 

Following Ref.~\cite{beloy2014atomic}, the effective temperature ($T^\mathrm{eff}$) that enters into the BBR shift calculation is such that \begin{equation}
    (T^\mathrm{eff})^4 = \sum_i \frac{\Omega^\mathrm{eff}_\mathrm{angle,i}}{4\pi} T_{i}^4,
\end{equation} where $T_{i}$ is the temperature of the $i$-th surface. We estimate an effective temperature of $T^\mathrm{eff} = 303(5)\,\mathrm{K}$.

\section{Clock state polarizabilities and BBR shifts \label{sec:abinitiopol}}

The electric dipole ($E1$) polarizabilities of vibrational states in the $X^1\Sigma_g^+$ potential are calculated using the sum-over-states approach \cite{bonin1997electric}. For states with total angular momentum $J=0$, the polarizability is a scalar quantity independent of the polarization of the electromagnetic field. Let $|i\rangle$ represent the rovibronic wavefunction of $X(v,J=0)$ and $|f\rangle$ represent the rovibronic wavefunction of a state that is $E1$-allowed from $|i\rangle$. The polarizability of a molecule in $X(v,J=0)$ is \begin{equation}\label{eq:polEqn1}
\alpha_v(\omega)=\frac{1}{\hbar}\sum_{f}|\langle f|d_0|i\rangle|^2\frac{2\omega_{fi}}{\omega_{fi}^2-\omega^2}.\end{equation} Here, $d_0$ is the component of the electric dipole moment operator parallel to the lab-frame quantization ($Z$) axis, $\omega_{fi}$ is the angular frequency of the rovibronic transition, and $\omega$ is the angular frequency of the electromagnetic field. The dc  (static) polarizability is recovered for $\omega = 0$.

\begin{figure}
\centering
\includegraphics[width=\columnwidth]{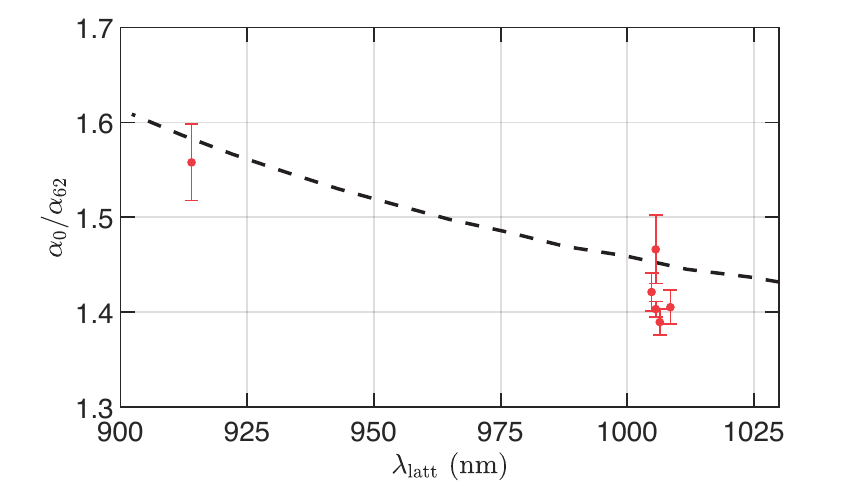}
\caption{The baseline ac polarizability ratio of the clock states in this study ($\alpha_0/\alpha_{62}$) at various trap laser wavelengths. Red circles are experimental measurements. The black dotted line is the theory calculation. $X\rightarrow (1)1_u$ resonances are excluded from this plot, but are included in the calculation. Error bars represent 1$\sigma$ statistical uncertainties. \label{fig:polexvsth}}
\end{figure}

The sum in Eq.~(\ref{eq:polEqn1}) is evaluated for bound-to-bound transitions to singlet ungerade excited potentials and bound-to-continuum transitions by discretizing the continuum. We include contributions from the $^1\Sigma^+_u$ potentials correlating to ${^1S}+{^1P}$ and ${^1S}+{^1D}$, as well as the $^1\Pi_u$ potentials correlating to ${^1S}+{^1P}$, ${^1S}+{^1D}$, ${^3P}+{^3P}$ and ${^3P}+{^3D}$. The $(1){^1}\Sigma_u^+$ potential is taken from the \textit{ab initio} calculation in Ref.~\cite{Skomorowski2012jcp}, while the doubly-excited $(3)^1\Pi_u$ and $(4)^1\Pi_u$ potentials were calculated using the multireference configuration interaction method (MRCI) with the MOLPRO package \cite{werner2019molpro}. The remaining potentials (including $X^1\Sigma_g^+$) are empirical \cite{Stein2010,Stein2011}. We omit spin-orbit and non-adiabatic couplings between the potentials. The convergence of our results is not changed by the inclusion of further high-lying potentials.

Laser wavelengths in the range 800--1200 nm can drive transitions from $X^1\Sigma_g^+$ to the short-range part of $(1)1_u$. These singlet-triplet transitions are relatively weaker than singlet-singlet ones, but become important if the laser is tuned near a resonance; e.g., in the case of a magic wavelength optical trap. To properly account for these situations, we additionally include the Morse/Long-range potential of $(1)1_u$ from Ref.~\cite{leung2020transition} in the polarizability sum. Figure~\ref{fig:polexvsth} shows the experimentally measured ac polarizability ratio $\alpha_0/\alpha_{62}$ over a range of wavelengths, determined using a frequency-only method \cite{leung2021ultracold,leung2020transition}, along with the calculation using Eq.~(\ref{eq:polEqn1}) showing consistency within $<20\%$.

To calculate the BBR shift, we use the formulas derived in Ref.~\cite{porsev2006multipolar} and properly adapt them for molecular states. In Ref.~\cite{porsev2006multipolar}, approximations were made to express the BBR shift in terms of the dc  polarizability and a power series in $k_BT^\mathrm{eff}/(\hbar \omega_{fi})$. These approximations are also valid for our case since $\omega_{fi}/(2\pi c) > 8000 \,\mathrm{cm}^{-1}$, corresponding to characteristic temperatures of $>$11500 K much greater than $T^\mathrm{eff}$. The so-called dynamic term contributes less than $0.5\%$ to the total BBR shift. The correction to the vibrational clock frequency at $T^\mathrm{eff}=303(5)\,\mathrm{K}$ is calculated to be $ -0.70(14)\,\mathrm{Hz}$. Here, we quote a conservative fractional uncertainty of $20\%$ based on the level of consistency between the measured and calculated ac polarizabilities [Fig.~\ref{fig:polexvsth}].

\section{Other clock systematics \label{sec:othersys}}

\textit{Effects of magnetic field}.---The use of singlet and irrotational $X^1\Sigma_g^+(J=0)$ clock states confer a high degree of insensitivity to external magnetic fields. Hyperfine sublevels are absent in our dimer assembled from $^{88}\mathrm{Sr}$, which has a total nuclear spin of zero. While the excited molecular states of $^{88}\mathrm{Sr}_2$ near the intercombination line have been thoroughly studied and modeled in previous work \cite{mcguyer2013nonadiabatic,mcguyer2015control}, an equivalent quantum chemistry calculation of higher-order Zeeman shifts of $X^1\Sigma_g^+(J=0)$ ground states is beyond the scope of this paper. Even without detailed theoretical modeling, we may experimentally investigate the extent to which our measurements are affected by magnetic field effects, including hypothetical Zeeman shifts of the clock states. We vary the applied magnetic field during clock interrogation with a lever arm of 3.2 G. The larger applied magnetic field slightly changes the photoassociation efficiency, which we partially compensate for by simultaneously varying the initial molecule number. Interleaved measurements obtain a differential shift of 0.05(41) Hz. For hypothetical shifts that scale quadratically with magnetic field strength, the measurement suggests that these contribute $<5\times10^{-16}$ to the systematic uncertainty.


\textit{dc  Stark shift}.---The stainless-steel vacuum chamber is electrically grounded, and the molecules are held at a distance of approximately 120 mm from each of the fused silica viewports. We have operated the vacuum system for over a decade. Thus, we expect any stray charges on the viewports to have migrated and decayed to a negligible amount. Even if a hypothetical, improbably large voltage difference of 20 V were present between two opposite-facing viewports, using the dc  polarizability difference computed with our model, the dc Stark shift multiplied by the number of such viewport pairs would amount to $<30\,\mathrm{mHz}$, or $<10^{-15}$ in fractional units.

\textit{Doppler shifts and phase chirps}.---First-order Doppler shifts result from the relative motion of the lattice anti-nodes and the phases of the probe lasers. For example, this may be due to the mechanical motion of the lattice retroreflector, or phase chirps arising from the pulsing of an acousto-optic modulator (AOM) that diffracts a probe beam. Our upleg clock laser is pulsed by AOM1 before delivery to the molecules. A common solution in lattice clocks is to perform fiber noise cancellation of the probe(s) using the lattice retroreflector either as the phase reference surface or as a rigid support for a separate surface \cite{falke2012delivering}, which we will implement in future work. If uncompensated, AOM phase chirps can result in shifts as large as $\sim$100 mHz. We do not study phase chirps in this work. Consequently, we quote a conservative upper bound of $10^{-14}$ for shifts originating from this effect. The second-order Doppler shift is $<10^{-19}$ for the typical thermal speed of our molecule.

\textit{Lattice tunneling}.---For the operational molecular trapping frequencies and temperature, we estimate that over $99\%$ of molecules occupy motional quantum numbers $n<8$, with $41\%$ in the ground motional band ($n=0$). A 1D lattice band structure calculation at the operational trap depth of $U_\mathrm{opt}= 487\, E_r$ involving 1000 lattice sites (or equivalently, 1000 Fourier components) indicates that the bandwidth of $n=7$ is $2\times 10^{-5}\, E_r$, which translates to 0.02 Hz for our molecular mass ($M$) and lattice wavelength ($\lambda_\mathrm{latt}$). As a check, we verified that our calculation quantitatively reproduces the results of identical band structure calculations in available lattice clock literature \cite{lemonde2005optical,falke2014strontium}. We thus quote an upper bound on Doppler-like shifts due to the delocalization of the molecular wavefunction to be $<1\times10^{-15}$.

\textit{Lattice light shift model}.---We fit quadratic polynomials to the measured clock shifts for the lattice light shift evaluation to account for the observed hyperpolarizability light shifts. Given our trap frequencies, sample temperatures, and the linear scaling of sample temperature with trap depth, the polynomial fit is a reasonable approximation. The $M1$-$E2$ shifts (that microscopically scale as $\sqrt{U_0}$ \cite{UshijimaKatori2018}) are included in the $\alpha^*$ effective parameter \cite{brown2017hyperpolarizability,beloy2020modeling,mcgrew2018atomic,bothwell2019jila}. In future work, calculating the differential $M1$ and $E2$ polarizabilities would help quantify the error associated with the thermal model. 

To test if higher-order polynomial terms are statistically significant, we fit the lattice light shifts to a cubic polynomial, with the quadratic and cubic coefficients as global fit parameters. The data suggests a cubic coefficient of $-1.2(9)\times 10^{-8}\,\mathrm{Hz}/E_r^3$. While the addition of a cubic term shifts the value of $f_\mathrm{zero}$, the estimated frequency correction, in this case, remains consistent with the applied correction in the main text within their uncertainties. As such, we limit our fitting to a quadratic polynomial for the present evaluation at the $10^{-14}$ level.

\textit{Line pulling}.---Under operational conditions, the radial trapping frequency is 311(2) Hz, which is $10\times$ larger than the full width at half maximum of the clock resonances. The clock and the lattice laser beams are coaligned over several meters, and the radial sidebands are not visible during normal clock operation. We model the radial sidebands as two Lorentzian peaks centered at their expected detuning from the carrier, with amplitudes equal to the size of the typical shot-to-shot signal variation. To put an upper bound on the line pulling effect, we compare the difference in the carrier line center returned by fitting a typical clock spectrum with the sum of three Lorentzians (i.e., two radial sidebands and one carrier), versus the case using just a single Lorentzian (as in Fig.~\ref{fig:expscheme}(c)). We estimate the line pulling error to be $<1\times10^{-15}$.

\textit{Scan-and-fit error}.---To estimate the effect of short-term cavity flicker noise on our peak fitting, we fit a linear function to a typical time series of molecular clock lines totaling $\sim 3000$ s (the typical duration for a single evaluation of a given systematic under interleaved clock operation). The magnitude of the linear coefficient is $<10\,\mathrm{mHz/s}$. For the present experiment, it takes $\sim$20 s to scan out all 15 points that make up the clock spectrum. Therefore, we estimate an upper bound for the scan-and-fit errors to be $<6\times 10^{-15}$.

As mentioned in Appendix~\ref{sec:ramanlaser}, the months-long average linear drift of the molecular clock line due to imperfect feedforward compensation is approximately $0.2 \,\mathrm{mHz/s}$. The feedforward parameters were set beforehand and unchanged during the campaign. This long-term drift would contribute a systematic offset of magnitude $1\times10^{-16}$, which is negligible for the current evaluation.

\textit{Gravitational redshift}.---We determine the elevation of our apparatus to be 51(5) m above mean sea level, which corresponds to a redshift correction of $-0.18(2)$ Hz. This correction has been added to give the reported absolute frequency of the clock.

\section{Timebase calibration \label{sec:timebase}}

\begin{figure*}
\centering
\includegraphics[width=2\columnwidth]{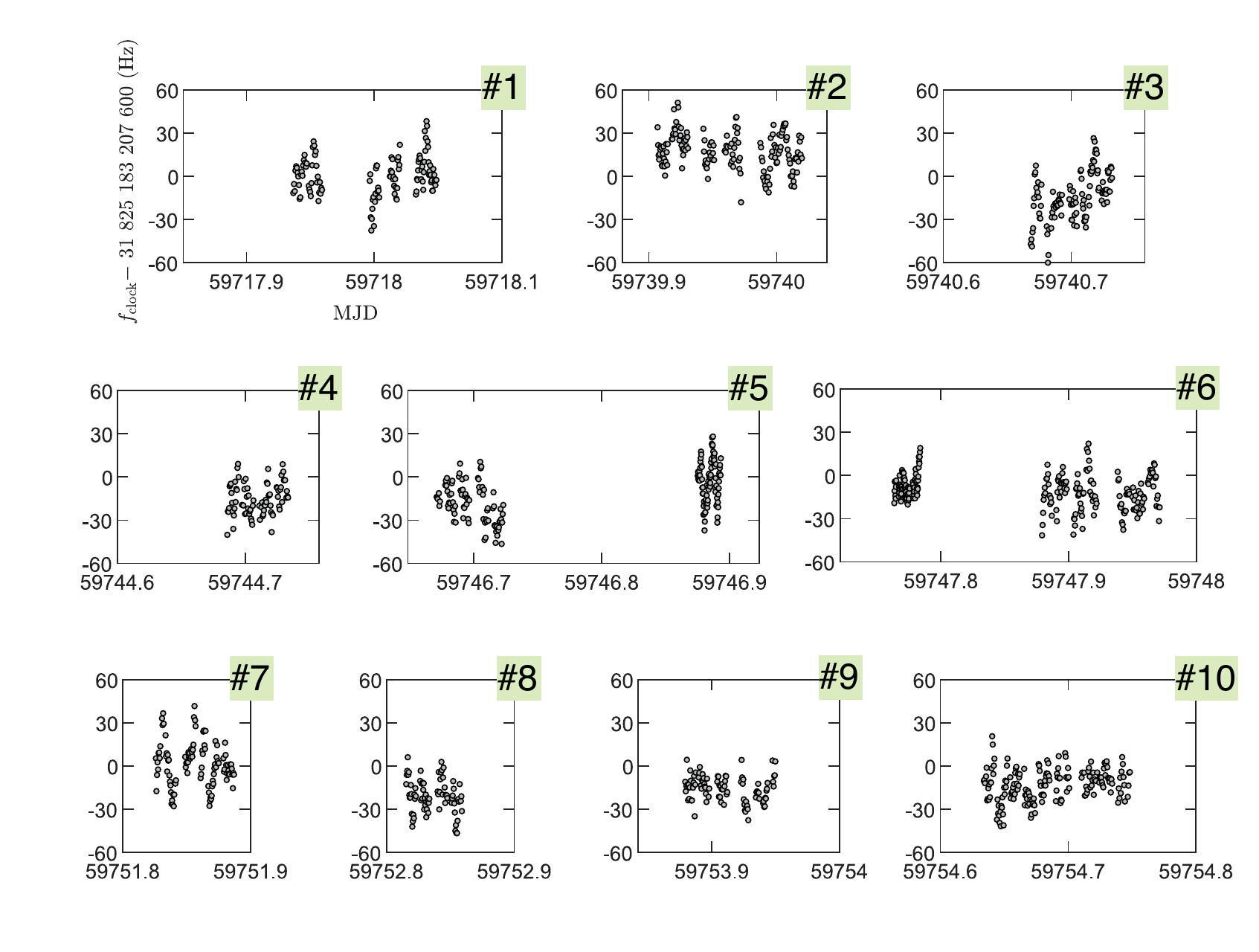}
\caption{Expanded dataset for the absolute clock frequency measurement. For experimental ease, the clock lasers and frequency comb are not actively steered toward the molecular resonance in real time. Instead, each measurement (filled grey circle) is derived from a single scan of the molecular clock transition under operational conditions, and the repetition rate of the frequency comb is counted with a zero dead time frequency counter (1 s gate time). The shot-to-shot noise in the counted repetition rates is lowered by taking their average over a short window of time centered on the timestamp of the corresponding molecular clock measurement. The data is plotted against the time of measurement (Modified Julian Date, MJD). The frequency values have been corrected for all systematic errors in Table~\ref{tab:systable}, the Rb clock calibration, and the gravitational redshift (Appendix~\ref{sec:othersys}). Error bars are not shown for visual clarity. Trial numbers on the top right corner of each plot correspond to those in Fig.~\ref{fig:absfreq}(a). The histogram of all measurements is shown in Fig.~\ref{fig:absfreq}(b). The axes labels are displayed in the plot for Trial 1 (top left), and are the same for the remaining plots.}
\label{fig:timeseries}
\end{figure*}

The frequency chain shown in Fig.~\ref{fig:expscheme}(b) can be algebraically written as \begin{align}
    \frac{f_\mathrm{clock}}{f_\mathrm{SI}} = &\frac{f_\mathrm{clock}}{f_\mathrm{Rb}} \times \frac{f_\mathrm{Rb}}{f_\mathrm{GPS}} \nonumber\\ &\times \frac{f_\mathrm{GPS}}{f_\mathrm{UTC(USNO)}} \times\frac{f_\mathrm{UTC(USNO)}}{f_\mathrm{TAI}}\times \frac{f_\mathrm{TAI}}{f_\mathrm{SI}}, \label{eq:freqchainCont}
\end{align} where $f_\mathrm{SI} \equiv 1\,\mathrm{Hz}$ is the SI unit of frequency such that $f_\mathrm{clock}/f_\mathrm{SI}$ is the numerical value of the absolute frequency of the molecular clock. $f_\mathrm{clock}/f_\mathrm{Rb}$ is the molecular clock frequency relative to the Rb clock that serves as our lab timebase (calculated using the molecular clock line centers, the counted repetition rate of the frequency comb, and the molecular clock systematic corrections). $f_\mathrm{Rb}/f_\mathrm{GPS}$ is the frequency of the Rb clock relative to GPS time. GPS time is closely steered toward UTC(USNO), and $f_\mathrm{GPS}/f_\mathrm{UTC(USNO)}$ is the frequency of GPS time relative to UTC(USNO). Finally, $f_\mathrm{UTC(USNO)}/f_\mathrm{TAI}$ is the frequency of UTC(USNO) relative to TAI, and $f_\mathrm{TAI}/f_\mathrm{SI}$ is the frequency of TAI relative to the SI second. 

The molecular clock is operated intermittently due to its complexity and the practical challenges of our present experimental apparatus. As such, the clock was not continuously phase-linked with the SI second. To address this, we expand Eq.~(\ref{eq:freqchainCont}) as \begin{align}
    \frac{f_\mathrm{clock}}{f_\mathrm{SI}} = &\frac{f_\mathrm{clock}}{f_\mathrm{Rb}(\mathcal{T}_1)} \times \frac{f_\mathrm{Rb}(\mathcal{T}_1)}{f_\mathrm{GPS}(\mathcal{T}_1)} \times \frac{f_\mathrm{GPS}(\mathcal{T}_1)}{f_\mathrm{GPS}(\mathcal{T}_2)} \nonumber\\ &\times \frac{f_\mathrm{GPS}(\mathcal{T}_2)}{f_\mathrm{UTC(USNO)}(\mathcal{T}_2)} \times\frac{f_\mathrm{UTC(USNO)}(\mathcal{T}_2)}{f_\mathrm{TAI}(\mathcal{T}_2)}\nonumber\\&\times \frac{f_\mathrm{TAI}(\mathcal{T}_2)}{f_\mathrm{SI}}.
\end{align} Here, $\mathcal{T}_1$ is the typical up time of the molecular clock corresponding to one measurement trial [Fig.~\ref{fig:timeseries}], $\mathcal{T}_2 = 1\,\mathrm{month}$ corresponds to the time period for the publication of the Circular T \cite{arias2011timescales}, and the bracketed time explicitly states the duration over which the given frequency is averaged. We have assumed that the SI second and molecular clock frequency are unchanging in time. 

Over the length of the campaign, the scale intervals of GPS time, UTC(USNO), and International Atomic Time (TAI) differed by $\lesssim 10^{-15}$, and the daily fractional changes in the frequency of GPS time relative to TAI are $\lesssim 10^{-14}$ \cite{circularT}. Therefore, for the present study, except for $f_\mathrm{Rb}(\mathcal{T}_1)/f_\mathrm{GPS}(\mathcal{T}_1)$ and $f_\mathrm{clock}/f_\mathrm{Rb}(\mathcal{T}_1)$, all other ratios contribute a negligible uncertainty and may be assumed to be unity. This includes the extrapolation ratio, $f_\mathrm{GPS}(\mathcal{T}_1)/f_\mathrm{GPS}(\mathcal{T}_2)$, which is the frequency of GPS time during the molecular clock up times versus the frequency of GPS time broadcasted by the constellation over a month. 

Each TIC measurement is started by the rising edge of the 1 PPS from the Rb clock and stopped by the rising edge of the 1 PPS from the GNSS receiver. Thus, the instantaneous fractional frequency offset of the Rb clock relative to the frequency of GPS time, $r= [f_\mathrm{Rb}/f_\mathrm{GPS}] -1$, is quantified by the instantaneous slope of the logged TIC measurements as a function of elapsed time. This measurement is susceptible to noise in the satellite link, as well as the instabilities of GPS time and the Rb clock. Comparisons with an identical, independent free-running Rb clock indicate that the Rb clock reaches an instability flicker floor of approximately $3\times10^{-13} \equiv \sigma_\mathrm{Rb}$ after $\sim 5\times10^3$ s of averaging time (comparable to typical durations of $\mathcal{T}_1$), but worsens to $\sim 10^{-12}$ for time periods over 24 hours. This poses a conundrum, because at least 24 hours of continuous averaging is typically required to achieve an inaccuracy and instability of $<10^{-13}$ using one-way GPS time transfer \cite{lombardi2001time}, but the Rb clock is not a good flywheel on these time scales. 

Rubidium microwave standards are more susceptible to unpredictable environmental perturbations than, for instance, hydrogen masers \cite{marlow2021review}, making it challenging to construct a reliable noise model. Therefore, for every trial, we operationally extract $r = [f_\mathrm{Rb}(\mathcal{T}_1)/f_\mathrm{GPS}(\mathcal{T}_1)] -1$ through linear fitting of the TIC measurements as a function of elapsed time, restricting the fits to the durations coinciding with the up time segments of the molecular clock. 

We judged a detailed characterization of the satellite link to be beyond the scope of this work. Geometric multipath effects and the diurnal variation in the ionosphere may introduce a systematic offset ($\approx 2\times 10^{-13}\equiv \sigma_\mathrm{GPS,sys}$), since a majority of the molecular clock up times were in the afternoon.

We estimate the fractional uncertainty of the extracted values of $r$ to be $\sqrt{\sigma_\mathrm{Rb}^2+\sigma_\mathrm{GPS,tot}^2}$, where $\sigma_\mathrm{GPS,tot}^2=\sigma_\mathrm{GPS,stat}^2+\sigma_\mathrm{GPS,sys}^2$ and $\sigma_\mathrm{GPS,stat} \approx 10^{-13} \times \sqrt{86400/\mathcal{T}_1\,[s]}$. The uncertainty from linear fitting is an order of magnitude smaller. Occasionally, we manually realigned the Rb clock frequency relative to GPS if it exceeded a fractional offset of $1\times10^{-11}$. This is not done during the molecular clock up times, nor within 24 hours of those segments to let the Rb clock settle. For every trial, we add a unique frequency correction $r\times [f_\mathrm{GPS}(\mathcal{T}_1)/f_\mathrm{GPS}(\mathcal{T}_2)] \times [f_\mathrm{clock}/f_\mathrm{Rb}(\mathcal{T}_1)]$ to $f_\mathrm{clock}/f_\mathrm{Rb}(\mathcal{T}_1)$, obtaining the absolute frequency of the molecular clock.

%

\end{document}